\documentclass[12pt]{article}
\usepackage{amsmath,amssymb}

\newcommand{\be}{\begin{equation}}
\newcommand{\ee}{\end{equation}}
\newcommand{\bea}{\begin{eqnarray}}
\newcommand{\eea}{\end{eqnarray}}
\newcommand{\beas}{\begin{eqnarray*}}
\newcommand{\eeas}{\end{eqnarray*}}
\newcommand{\bi}{\begin{itemize}}
\newcommand{\ei}{\end{itemize}}
\newcommand{\bc}{\begin{center}}
\newcommand{\ec}{\end{center}}
\newcommand{\bfl}{\begin{flushleft}}
\newcommand{\efl}{\end{flushleft}}
\newcommand{\bfr}{\begin{flushright}}
\newcommand{\efr}{\end{flushright}}

\def\6{\partial}

\begin{document}

\title{Local Physical Coordinates from Symplectic Projector Method}

\author{M. A. De Andrade$^{a}$\footnote{E-mail: marco@gft.ucp.br} , ~ 
M. A. Santos$^{b}$\footnote{E-mail: masantos@gft.ucp.br} ~ and ~ I. V. Vancea$^{c}$
\footnote{E-mail: ivancea@unesp.ift.br, vancea@cbpf.br. On leave from Babes Bolyai 
University of Cluj, Romania.}}

\date{\small $^a$Grupo de F\'{\i}sica Te\'{o}rica, Universidade 
Cat\'{o}lica de Petr\'{o}polis \\ 
Rua Bar\~ao de Amazonas 124, 25685-070,  Petr\'opolis - RJ, Brasil \\ and \\
Departamento de F\'{\i}sica, Universidade Federal de Juiz de Fora \\ 
Cidade Universit\'{a}ria, Caixa Postal 656, 36036-330, Juiz de Fora - MG, Brasil. \\
\vspace{.5cm}
$^b$Departamento de F\'{\i}sica, Universidade Federal Rural do Rio de Janeiro\\
23851-180, Serop\'{e}dica - RJ, Brasil\\ and \\
Centro Brasileiro de Pesquisas F\'{\i}sicas,\\
Rua Dr. Xavier Sigaud 150, 22290-180 Rio de Janeiro - RJ, Brasil\\
\vspace{.5cm} 
$^c$Instituto de F\'{\i}sica Te\'{o}rica , Universidade Estadual Paulista \\ 
Rua Pamplona 145, 015405-900.  S\~ao Paulo - SP, Brasil \\}

\maketitle

\abstract{The basic arguments underlying the symplectic projector method are
presented. By this method, local free coordinates on the constraint surface
can be obtained for a broader class of constrained systems. Some interesting
examples are analyzed.}

\section{Introduction}

Over the last three decades, understanding the constrained systems has been
a fundamental problem in theoretical physics. The reason for that lies in
the principle of gauge invariance and the principle of covariance of quantum
field theory and gravity. It follows from these principles that the models
of the fundamental interactions should be formulated in terms of fields that
are subjected to constraints. The most effective method to deal with
constrained system is the BRST method, in which the classical and quantum
quantities are found as objects in the cohomology group of a nilpotent
operator which encodes the symmetry of the system. The BRST operator acts on
functionals defined on an \emph{enlarged} configuration (or phase) space, in
which new variables (of opposite Grassmann parity) are added to original
ones. ``The physical surface'' is embedded in this space and the physics on
it is retrieved by solving the cohomology problem. Although this method, in
its various formulations, deals with all known situations in a remarkable
efficient way, allowing one to solve complex problems as the identification
of anomalies, Schwinger terms and renormalization, one looses the intuitive
picture of the physical space (see, for a review of BRST \cite{ht}.)
However, it is possible to work with the constrained systems by \emph{%
reducing} the space to the physical subspace instead of enlarging it.

The aim of this letter is to review a method for finding the local
coordinates on the constraint surface on which the dynamics takes place. The
original idea was to construct a projector in the configuration space that
projects the global coordinates of the space onto the constraint surface.
Thus, free local coordinates can be obtained on the surface and the local
dynamics of the system can be written entirely in terms of these
coordinates. This method was initially developed for holonomic constraints
in the configuration space \cite{cma}, extended to phase space
\cite{pp}, and tested in the analyses of gauge field theories 
\cite{qe,cl,mj,alv,mmi1}. Despite the fact that it is
less powerful than BRST method and its applicability is quite restricted, it
might be simpler to handle in various situations when one is not interested
in maintaining the full generality of the BRST or when one deals with
collective processes in many-body systems. Results along this line have been
obtained for quantum electrodynamics \cite{qe}, Chern-Simons in three
dimensions \cite{alv}, and non-commutative strings \cite{mmi1}
among others, where physical coordinates that simplifies the analysis of
these systems have been given

The paper is organized as follows. In Section 2 we are going to present the
construction of projector for holonomic constraints in the configuration
space following \cite{cma}. The projected variables and the symplectic
projector are discussed in Section 3. In Section 4 we analyze two examples:
the electrodynamics \cite{qe} and a toy model recently used in discussing
the properties of strings in the presence of a $B$-field \cite{suss}. The
last section is devoted to discussions.

\section{Projector on the constraint surface}

Consider a physical system described by a set of coordinates $x^i$, $%
i=1,2,\ldots,n$ on a manifold $E^n$ which is called the \emph{configuration
space}. The classical motion of the system is given by a curve $\mathbf{x}%
(t) $ in $E^n$, where for a non-relativistic system $t$ is the time. This
curve is obtained by minimizing a functional action which is the time
integral of a Lagrangian 
\begin{equation}
S[\mathbf{x}(t)] = \int dt~ L[\mathbf{x}(t), \dot{\mathbf{x}}(t)],
\label{actionfunc}
\end{equation}
where more general Lagrangians on higher derivatives of $\mathbf{x}(t)$ and
explicit dependence of $t$ can be considered in (\ref{actionfunc}). From the
geometric point of view, the Lagrangian functional in (\ref{actionfunc}) is
a function on the tangent bundle $T(E^n)$ that takes any pair $(\mathbf{x},%
\mathbf{v})$ into a real number 
\begin{equation}
L : T(E^n)~\rightarrow~R.  \label{tblangr}
\end{equation}
On the trajectory curve the vector $\mathbf{v}(t)=\dot{\mathbf{x}}(t)$. The
trajectory is the solution of the second order differential equations
(Euler-Lagrange) that can be obtained from (\ref{actionfunc}). In order to
classify these trajectories, one has to transform the Euler-Lagrange
equations into first order differential equations. This can be done by
introducing new variables $p^i$ from the cotangent bundle $T^{*}(E^n)$ to
the configuration space. The trajectory is now a curve in $T^*(E^n)$ and the
equations can be transformed into first order differential equations
(Hamilton) written as usual in terms of the Hamiltonian which is a real
functional in the cotangent bundle $H(\mathbf{x}, \mathbf{p})$ or in the 
\emph{phase space}. However, the Hamiltonian and the solutions of the
equations of motion may not exist for every point of the configuration
space. This happens when the following Hessian is degenerate 
\begin{equation}
W_{ij}\equiv \frac{\partial ^{2}L}{\partial {\dot{x}}_{i} \partial {\dot{x}}%
_{j}}.  \label{m2}
\end{equation}
Nevertheless, the solutions of the equations of motion and the Hamiltonian
always exists on a submanifold of the configuration space or phase space
given by the momentum map 
\begin{equation}
p_i = \frac{\partial L(\mathbf{x},\mathbf{p})}{\partial v^i},  \label{mommap}
\end{equation}
which can also be expressed analytically by a set of equations called \emph{%
primary constraints} 
\begin{equation}
\psi_{\mu}(\mathbf{x},\mathbf{p}) =0.  \label{primary}
\end{equation}
The dynamics should hold on this surface. If one divides the constraints in
a subset $\chi_a$ that satisfies the relations 
\begin{equation}
\{ \chi_a, \psi_{\mu} \} \approx 0,  \label{firstclasscon}
\end{equation}
called \emph{first class constraints} and the remaining set of relations $%
\phi_m$ called the \emph{second class constraints}, then the matrix 
\begin{equation}
\Delta_{mn} = \{\phi_m, \phi_n \}  \label{submatrix}
\end{equation}
is invertible and fixes the canonical Hamiltonian. If one has a theory with
only second class constraints the dynamics of any observable (i.e. a phase
space function $A\left(\mathbf{x},\mathbf{p}\right)$) is given by the
following relation 
\begin{equation}
\dot{A}\approx \left\{ A,H_{c}\right\} -\left\{ A,\phi _{m},\right\} \Delta
_{mn}^{-1}\left\{ \phi _{n},H_{c}\right\}.  \label{m4}
\end{equation}
Here, $\Delta _{mn}^{-1}$ is the inverse of the ( non-singular) matrix $%
\Delta _{mn}$. In terms of the Dirac brackets, we can rewrite (\ref{m4}) as 
\begin{equation}
\dot{A}\approx \left\{ A,H_{c}\right\} _{D}.  \label{m6}
\end{equation}
The first class constraints generate gauge transformations and one has to
fix the gauge before applying the formula (\ref{m6}).

It was originally proposed by Dirac to make use of the Dirac brackets in
passing to quantum commutators. However, there is a drawback of the Dirac
brackets structure, namely that, as extended brackets, the commutators
obtained from them loose the Kronecker delta structure. One way to makeshift
around this problem is to maintain the canonical bracket structure and to
reduce the analysis of the system to the physical subspace which is embedded
in the configuration (phase) space. This goal can be achieved by
constructing the projection operators onto the physical surface. We are
going to present a method for doing that in the configuration space of a
discrete system. The extension to the phase space is obvious since the
geometric structures underlying the construction are the same. The
presentation follows the line of the original paper \cite{cma}.

Let us consider a discrete system for which the configuration space $E^n$ is
taken to be Euclidean and isomorphic to $R^n$. The global Cartesian
coordinates are denoted by $x^i$, where $i=1,\ldots , n$. We assume that the
physical system is subject to holonomic constraints in the configuration
space. The constraint surface is described by the following set of
independent equations 
\begin{equation}
\Sigma ~:~~~ \phi^I (x^i)=0,  \label{ion-consurf}
\end{equation}
where $I=1,2,\ldots, r$ label the relations among coordinates. The surface $%
\Sigma$ is geometric and independent of the dynamics. Therefore, the
geometrical objects related to it are invariant under the evolution of the
system. However, the dynamics takes place on $\Sigma$ and we want to find a
local coordinate system on it which is free of constraints.

The first point to be noted is that one can construct a local coordinate
system at every point $x \in E^n$ of global coordinates $x^i$ if a regular
matrix field $M(x^i)$ is given. Since the construction is purely geometric,
the elements of $M(x^i)$ should be time independent. If $\mathbf{e}_i$ is
the orthogonal basis of $E^n$ associated to the global Cartesian
coordinates, the local basis $\mathbf{f}_a(x)$, where $a=1,2,\ldots,n$
associated to the coordinate system in $P$ is given by 
\begin{equation}
\mathbf{f}_a(x) = M^{i}_{a}(x)\mathbf{e}_i .  \label{ion-localbasis}
\end{equation}
In general, $M^i_a(x)$ will define some local coordinates on $\Sigma$, too.
The configuration of the system at a moment $t$ will be given by a Cartesian
vector $\mathbf{x}(t)$ in the configuration space that points to a point on
the trajectory embedded in $\Sigma$.

In order to be able to project onto the constraint surface, one picks up a
specific local basis at each point $x \in \Sigma$ by splitting the tangent
space $T^n_x(E)$ to $E^n$ at $x$ into the direct sum $T^{n-r}_x(\Sigma)
\oplus N^r_x(\Sigma)$. Here, $T^{n-r}_x(\Sigma)$ is the tangent space to $%
\Sigma$ at $x$ and $N^r_x(\Sigma)$ is the normal space. The local basis is
thus split as $\mathbf{f_a}(x) = \{ \mathbf{n}_I(x), \mathbf{t}_{\alpha }(x)
\}$, where $\mathbf{n}_I(x) $ form the normalized basis of $N^r_x(\Sigma)$
while $\mathbf{t}_{\alpha }$, $\alpha = 1,2,\ldots,n-r$ stands for a basis
in the tangent space $T^{n-r}_x(\Sigma)$. The components of $\mathbf{n}_I(x)$
are given in terms of the derivatives of $\phi^I (x)$ 
\begin{equation}
\mathbf{n}_{Ii} (x) = \frac{\partial_i \phi^I(x)}{| \partial_i \phi^I(x) |},
\label{ion-comp}
\end{equation}
where we have used the Euclidean product in $E^n$ to compute the norm of the
vector. Since the normal vectors $\partial_i\phi^I(x)$ are not necessarily
orthogonal, they generate a metric $g^{IJ}(x)$.

With these objects at hand it is easy to construct a projector that projects
the vectors of $E^n$ in the neighborhood of $x$. One way to see that is by
introducing the canonical basis $\mathbf{n}^{*I}(x)$ in the cotangent space $%
T^{*(n-r)}_x$ associated to $\{\mathbf{n}_I(x) \}$ in terms of which the
sought for projector $\Lambda (x)$ can be written as 
\begin{equation}
\Lambda (x) = \mathbf{1} - \sum_{I=1}^r \mathbf{n}_I(x)\mathbf{n}^{*I}(x).
\label{ion-project}
\end{equation}
The projector (\ref{ion-project}) will pick up the parallel components to
the tangent space to $\Sigma$ at $x$ of any vector of the configuration
space. As was pointed out in \cite{cma}, the local tangent space is free of
constraints. In order to find out the local configuration of the system at
any moment $t$ one has to project the configuration vector $\mathbf{x}(t)$
by acting with $\Lambda$ on it. The two spaces $T^{n-r}_x$ and $N^r_x$ are
given by the eigenvectors of the complement projector $Q(x)=\mathbf{1}-
\Lambda(x)$ associated to the eigenvalues 0 and 1, respectively.

By picking up an appropriate matrix $M^i_a(x)$, one can split up the
components of the configuration vector into two sets $\{ x^{\alpha}(t),
x^I(t)\}$ that satisfy the following boundary conditions \cite{cma} 
\begin{equation}
\dot{x}^{I}(t)=0~~~,~~~x^{I}(t) = \phi^{I}(x)=0,  \label{ion-bc}
\end{equation}
where the vectors have been ordered as follows $I =n-r+1,\ldots,n$. The
general regular transformation to the local coordinates that satisfy the
relations (\ref{ion-bc}) leaves the Lagrangian invariant (but not the
Hamiltonian) and therefore the unconstrained Lagrangian can be locally
written as 
\begin{equation}
L^{\prime}(x^{\alpha}(t),{\dot{x}}^{\alpha}(t)) = L(x^i(t), {\dot{x}}^{i}(t))
\label{ion-lagrange}
\end{equation}
The local dynamics is given by free Euler-Lagrange equations obtained from $%
L^{\prime}$ in terms of $x^{\alpha }$ only, or from $L$ with the boundary
conditions (\ref{ion-bc}).

This construction extends easily to the phase space. In the case of the
quantum field theory it can be reproduced formally in the same way, however
some care should be taken when the corresponding geometrical objects are
defined. It is important to emphasize once again that the full construction is
local on the constraint surface. Finding the global properties of the
projector is an interesting issue which has not been addressed yet.

\section{Symplectic $\Lambda$ projector and physical variables}

Let us consider the phase space of a system with second class constraints $%
\phi_m$. The phase space coordinates are denoted by $\xi^M$, where $%
M=1,2,\ldots,2n$. The projector defined in (\ref{ion-project}) has the
following form 
\begin{equation}
\Lambda ^{MN} = \;\; \delta ^{\,MN}- J^{ML}\,\frac{\delta{\phi}_{m}}{\delta
\xi ^{L}}\,\Delta^{-1}_{mn}\, \frac{\delta \phi _{n}} {\delta \xi^{N}}.
\label{m7}
\end{equation}
The projector $\Lambda^{MN}$ acts on vector coordinates ${\xi }^M$ of the
space-phase endowed with a symplectic two-form $J^{MN}$. In theories with
gauge symmetry, the gauge must be fixed firstly in order to apply this
method. The local \emph{physical} or \emph{true} variables $\mathbf{\xi }%
^{*} $ are obtained as a result of the action of (\ref{m7}) on the vector $%
\mathbf{\xi :}$ 
\begin{equation}
\xi ^{*M}=\Lambda ^{MN}\xi ^{N}.  \label{m8}
\end{equation}
Starting with a $2n$-dimensional phase-space in the presence of $2m$
second-class constraints, we are lead to a vector with $2(n-m)$ independent
components. When one projects the configuration vector, the coordinates
given by (\ref{m8}) encode the dynamics of the system as discussed in the
previous section. In order to write down the Hamiltonian that determines the
dynamics one has to find out the boundary conditions for the canonical
momenta associated with the coordinates that satisfy (\ref{ion-bc}). They
are found to be given by the following relations \cite{cma} 
\begin{equation}
p_I(t)={\dot{p}}_I(t)=0.  \label{ion-bcmom}
\end{equation}
From (\ref{ion-bc}) and (\ref{ion-bcmom}) one can see that the Hamiltonian
is given by the original one written in terms of the coordinates that are
obtained after projection, i. e. the local coordinates on the physical
surface. These variables are independent, free of constraints and they obey
canonical commutation relations. The equations of motion follow from the
usual Hamilton-Jacobi equations: 
\begin{equation}
\stackrel{.}{\mathbf{\xi }}^{*}=\left\{ \mathbf{\xi }^{*},H^{*}\right\}
\label{m9}
\end{equation}

One notes that there is a structural similarity between the expression in (%
\ref{m7}) and the well-know fundamental Dirac brackets matrix\cite{dirac}: 
\begin{equation}
D^{MN}=\{\xi ^{M}\;,\;\xi ^{N}\}_{D}=J^{MN}-J^{ML}J^{KN}\,\frac{\delta \phi
_{m}}{\delta \xi ^{L}}\,\Delta _{mn}^{-1}\,\frac{\delta \phi _{n}}{\delta
\xi ^{K}}.  \label{m10}
\end{equation}
It is easy to see that the following relation relating the geometric
projection and the algebraic Dirac matrix holds: 
\begin{equation}
\Lambda =-DJ.  \label{m11}
\end{equation}
This simple relation connects two objects that have been obtained through
different constructions.

Let us make two remarks. Firstly, observe that if one computes the trace 
of the 
$\Lambda$-matrix, one obtains the number of the degrees of freedom of the
system. Secondly, the observables of the theory depend locally on the 
coordinates $\xi^*$ on
the constraint surface. Indeed, as for the Lagrangian, one has to take into
account the boundary conditions (\ref{ion-bc}) and (\ref{ion-bcmom}) in
determining any observables. But these relations actually express the
independence of the corresponding function on the local normal coordinates.

\section{Examples}

In this section we are going to illustrate the method of projectors in two
examples: the electrodynamics and a toy model used in the study of the
noncommutative D-branes. The first example has been studied in \cite{qe}.
The second one is presented here for the first time.

\subsection{Electrodynamics}

Let us apply the symplectic projector method to the Maxwell's electrodynamics
in the radiation gauge. Our purpose is to obtain the physical variables and
the Hamiltonian as was discussed above. The starting point is the canonical
Hamiltonian
\begin{equation}
\mathcal{H}=\int d^{3}x\left\{ \frac{1}{2}(\mbox{\boldmath$\pi$} ^{2}+\mbox{\boldmath$B$}^{2})+\pi _{\psi
}\gamma ^{0}(\mbox{\boldmath${\gamma \cdot \partial}$}-ie\,\mbox{\boldmath$\gamma\cdot{A}$}-im)\psi \right\}
\label{m12}
\end{equation}
together with the set of (second-class) constraints 
\begin{eqnarray}
\phi _{1} &=&\pi _{0} \\
\phi _{2} &=&\mbox{\boldmath$\partial\cdot\pi$ }+ie\,\pi _{\psi }\psi   \nonumber \\
\phi _{3} &=&A_{0}  \nonumber \\
\phi _{4} &=&\mbox{\boldmath$\partial\cdot{A}$}  \label{m13}
\end{eqnarray}
These are the informations we need to apply the procedure: we have the local
metric 
\begin{equation}
\Delta ^{-1}=\left( 
\begin{array}{llll}
0 & 0 & \delta ^{3}(x-y) & 0 \\ 
0 & 0 & 0 & \nabla ^{-2} \\ 
-\delta ^{3}(x-y) & 0 & 0 & 0 \\ 
0 & -\nabla ^{-2} & 0 & 0
\end{array}
\right) \;.  \label{m14}
\end{equation}
The matrix elements of the symplectic projector are given by the
following relation 
\[
\Lambda _{\nu }^{\mu }(x,y)=\delta _{\nu }^{\mu }\,\delta
^{3}(x-y)-\varepsilon ^{\mu \alpha }\int d^{\,3}\!\rho \;d^{\,3}\!\sigma
\,\delta _{\alpha (x)}\phi ^{i}(\rho )\,\Delta _{ij}^{-1}(\rho ,\sigma
)\,\delta _{\nu (y)}\phi ^{j}(\sigma )\;,
\]
where $\delta _{\alpha (x)}\phi ^{i}(\rho )\equiv \frac{
\delta \phi ^{i}(\rho )}{\delta \xi ^{\alpha }(x)}$ and $\left(
A_{0},A_{1},A_{2},A_{3},\psi ,\pi _{0},\pi _{1},\pi _{2},\pi _{3},\pi _{\psi
}\right) \longleftrightarrow \left( \xi _{1},\cdots ,\xi _{10}\right) $.
A simple algebra gives the $\Lambda $-matrix 
\begin{eqnarray}
&&\!\!\!\!\!\!\!\!\!\!\!\!\Lambda (x,y)=  \nonumber \\
&&{\!\!\!\!\!\!\!\!\!\!\!\!\left( 
\begin{array}{cccccccccc}
0\!\! & \!\!0\!\! & \!\!0\!\! & \!\!0\!\! & \!\!0\!\! & \!\!0\!\! & \!\!0\!\!
& \!\!0\!\! & \!\!0\!\! & \!\!0 \\ 
0\!\! & \!\!{\scriptstyle\delta ^{3}(x-y)-}\frac{\partial _{1}^{x}\partial
_{1}^{y}}{\nabla ^{2}}\!\! & \!\!-\frac{\partial _{1}^{x}\partial _{2}^{y}}{%
\nabla ^{2}}\!\! & \!\!-\frac{\partial _{1}^{x}\partial _{3}^{y}}{\nabla ^{2}%
}\!\! & \!\!0\!\! & \!\!0\!\! & \!\!0\!\! & \!\!0\!\! & \!\!0\!\! & \!\!0 \\ 
0\!\! & \!\!-\frac{\partial _{2}^{x}\partial _{1}^{y}}{\nabla ^{2}}\!\! & 
\!\!{\scriptstyle\delta ^{3}(x-y)-}\frac{\partial _{2}^{x}\partial _{2}^{y}}{%
\nabla ^{2}}\!\! & \!\!-\frac{\partial _{2}^{x}\partial _{3}^{y}}{\nabla ^{2}%
}\!\! & \!\!0\!\! & \!\!0\!\! & \!\!0\!\! & \!\!0\!\! & \!\!0\!\! & \!\!0 \\ 
0\!\! & \!\!-\frac{\partial _{3}^{x}\partial _{1}^{y}}{\nabla ^{2}}\!\! & 
\!\!-\frac{\partial _{3}^{x}\partial _{2}^{y}}{\nabla ^{2}}\!\! & \!\!{%
\scriptstyle\delta ^{3}(x-y)-}\frac{\partial _{3}^{x}\partial _{3}^{y}}{%
\nabla ^{2}}\!\! & \!\!0\!\! & \!\!0\!\! & \!\!0\!\! & \!\!0\!\! & \!\!0\!\!
& \!\!0 \\ 
0\!\! & \!\!\frac{-ie\,\xi _{5}(x)\partial _{1}^{y}}{\nabla ^{2}}\!\! & \!\!%
\frac{-ie\,\xi _{5}(x)\partial _{2}^{y}}{\nabla ^{2}}\!\! & \!\!\frac{%
-ie\,\xi _{5}(x)\partial _{3}^{y}}{\nabla ^{2}}\!\! & \!\!{\scriptstyle%
\delta ^{3}(x-y)}\!\! & \!\!0\!\! & \!\!0\!\! & \!\!0\!\! & \!\!0\!\! & \!\!0
\\ 
0\!\! & \!\!0\!\! & \!\!0\!\! & \!\!0\!\! & \!\!0\!\! & \!\!0\!\! & \!\!0\!\!
& \!\!0\!\! & \!\!0\!\! & \!\!0 \\ 
0\!\! & \!\!0\!\! & \!\!0\!\! & \!\!0\!\! & \!\!\frac{ie\partial _{1}^{x}\xi
_{10}(y)}{\nabla ^{2}}\!\! & \!\!0\!\! & \!\!{\scriptstyle\delta ^{3}(x-y)-}%
\frac{\partial _{1}^{x}\partial _{1}^{y}}{\nabla ^{2}}\!\! & \!\!-\frac{%
\partial _{1}^{x}\partial _{2}^{y}}{\nabla ^{2}}\!\! & \!\!-\frac{\partial
_{1}^{x}\partial _{3}^{y}}{\nabla ^{2}}\!\! & \!\!\frac{-ie\partial
_{1}^{x}\xi _{5}(y)}{\nabla ^{2}} \\ 
0\!\! & \!\!0\!\! & \!\!0\!\! & \!\!0\!\! & \!\!\frac{ie\partial _{2}^{x}\xi
_{10}(y)}{\nabla ^{2}}\!\! & \!\!0\!\! & \!\!-\frac{\partial
_{2}^{x}\partial _{1}^{y}}{\nabla ^{2}}\!\! & \!\!{\scriptstyle\delta
^{3}(x-y)-}\frac{\partial _{2}^{x}\partial _{2}^{y}}{\nabla ^{2}}\!\! & \!\!-%
\frac{\partial _{2}^{x}\partial _{3}^{y}}{\nabla ^{2}}\!\! & \!\!\frac{%
-ie\partial _{2}^{x}\xi _{5}(y)}{\nabla ^{2}} \\ 
0\!\! & \!\!0\!\! & \!\!0\!\! & \!\!0\!\! & \!\!\frac{ie\partial _{3}^{x}\xi
_{10}(y)}{\nabla ^{2}}\!\! & \!\!0\!\! & \!\!-\frac{\partial
_{3}^{x}\partial _{1}^{y}}{\nabla ^{2}}\!\! & \!\!-\frac{\partial
_{3}^{x}\partial _{2}^{y}}{\nabla ^{2}}\!\! & \!\!{\scriptstyle\delta
^{3}(x-y)-}\frac{\partial _{3}^{x}\partial _{3}^{y}}{\nabla ^{2}}\!\! & \!\!%
\frac{-ie\partial _{3}^{x}\xi _{5}(y)}{\nabla ^{2}} \\ 
0\!\! & \!\!\frac{ie\,\xi _{10}(x)\partial _{1}^{y}}{\nabla ^{2}}\!\! & \!\!%
\frac{ie\,\xi _{10}(x)\partial _{2}^{y}}{\nabla ^{2}}\!\! & \!\!\frac{%
ie\,\xi _{10}(x)\partial _{3}^{y}}{\nabla ^{2}}\!\! & \!\!0\!\! & \!\!0\!\!
& \!\!0\!\! & \!\!0\!\! & \!\!0\!\! & \!\!{\scriptstyle\delta ^{3}(x-y)}
\end{array}
\right) }.  \nonumber \\
&&
\end{eqnarray}
By projecting the symplectic vector we get the following physical coordinates 
\begin{equation}
\xi ^{1*}\left( x\right) =\xi ^{6*}\left( x\right) =0  \label{m16a}
\end{equation}
\begin{equation}
\xi ^{n*}\left( x\right) =A_{n-1}^{\perp }\left( x\right) ,n=2,3,4
\label{m16b}
\end{equation}
\begin{equation}
\xi ^{5*}\left( x\right) =\psi \left( x\right)   \label{m16c}
\end{equation}
\begin{equation}
\xi ^{10*}\left( x\right) =\pi _{\psi }\left( x\right)   \label{m16d}
\end{equation}
\begin{equation}
\xi ^{m*}(x)=\pi _{m-6}^{\perp }(x)+2ie\int d^{\,3}y\,\partial
_{m-6}^{x}\nabla ^{-2}\pi _{\psi }(y)\,\psi (y)\;,m=7,8,9\;.  \label{m16e}
\end{equation}
With these coordinates, we obtain the well-known Hamiltonian of 
electrodynamics \cite{sak} 
\begin{eqnarray}
\mathcal{H}^{*}&=&\int d^{\,3}x\,\frac{1}{2}\left(\mbox{\boldmath$\pi$}^{\perp 2}+\mbox{\boldmath$B$}^{2}\right)
+\pi _{\psi}\gamma^{0}\left(\mbox{\boldmath$\gamma\cdot\partial$}-ie\,\mbox{%
\boldmath$\gamma\cdot{A}$}^{\perp}-im\right)\psi \\
&&+\frac{e^{2}}{2\pi}\int d^{\,3}x\,d^{\,3}y\,\pi _{\psi
}(x)\,\psi(x)\,\nabla ^{-2}\pi _{\psi}(y)\,\psi(y)\;.  \label{m17}
\end{eqnarray}
The electrodynamics represents a basic example of how the symplectic projector 
method should be applied in the case of abelian gauge fields. 

\subsection{Toy Model}

One important class of constraints are the self-dual constraints intimately
related to the chiral symmetry. In \cite{suss} it was proposed a toy model
that exhibits such of constraints as a result of using noncommutative
coordinates. This models has been used in \cite{suss,Kim-Oh} to illustrate
some features of noncommutative open strings and D-branes in the Dirac
quantization.

The starting point is the following Lagrangian 
\begin{equation}
L=-\frac{eB}{2c}\dot{x}^{i}\varepsilon _{ij}\,x^{j}+e\Phi (x),  \label{10}
\end{equation}
which describes a charged particle interacting with an electro-magnetic
field of static potential $\Phi (x)$ and constant, large magnetic field $B$
in the limit when the kinetic energy term can be neglected or 
$m \rightarrow 0$.
The phase space of the model is 4-dimensional and its coordinates are
subjected to two (self-dual) constraints of second class 
\begin{equation}
\phi _{i}\equiv p_{i}+\frac{eB}{2c}\;\varepsilon _{ij}\,x^{j}\approx 0.
\label{11}
\end{equation}
Let us introduced the phase space vector $\mathbf{\xi }$ with the following
components 
\begin{equation}
\left( \xi ^{1},\xi ^{2},\xi ^{3},\xi ^{4}\right) \equiv \left(
x^{1},x^{2},p_{1},p_{2}\right) .  \label{12}
\end{equation}
The Dirac brackets for the coordinates are given by the following relations 
\begin{eqnarray}
\{x^{i}\;,\;x^{j}\}_{D} &=&\frac{c}{eB}\,\varepsilon ^{ij}  \label{13} \\
\{x^{i}\;,\;p_{j}\}_{D} &=&\frac{1}{2}\delta _{j}^{\,i}  \label{14} \\
\{p_{i}\;,\;p_{j}\}_{D} &=&-\frac{eB}{4c}\,\varepsilon _{ij}\;.  \label{15}
\end{eqnarray}
It is easy to see the the $D$-matrix is 
\begin{equation}
D={\left( 
\begin{array}{cccc}
0 & -\frac{c}{eB} & \frac{1}{2} & 0 \\ 
\frac{c}{eB} & 0 & 0 & \frac{1}{2} \\ 
-\frac{1}{2} & 0 & 0 & -\frac{eB}{4c} \\ 
0 & -\frac{1}{2} & \frac{eB}{4c} & 0
\end{array}
\right) }.  \label{16}
\end{equation}
Using the relation (\ref{m11}) we can write down the symplectic projector, 
\begin{equation}
\Lambda ={\left( 
\begin{array}{cccc}
\frac{1}{2} & 0 & 0 & \frac{c}{eB} \\ 
0 & \frac{1}{2} & -\frac{c}{eB} & 0 \\ 
0 & -\frac{eB}{4c} & \frac{1}{2} & 0 \\ 
\frac{eB}{4c} & 0 & 0 & \frac{1}{2}
\end{array}
\right) }\;.  \label{proj17}
\end{equation}
Finally, from (\ref{proj17}) we obtain the physical variables: 
\begin{equation}
\xi ^{\ast 1}=\frac{1}{2}\,\xi ^{1}+\frac{c}{eB}\,\xi ^{4}  \label{18}
\end{equation}
\begin{equation}
\xi ^{\ast 2}=\frac{1}{2}\,\xi ^{2}-\frac{c}{eB}\,\xi ^{3}  \label{19}
\end{equation}
\begin{equation}
\xi ^{\ast 3}=-\frac{eB}{4c}\,\xi ^{2}+\frac{1}{2}\,\xi ^{3}=-\frac{eB}{2c}%
\,\xi ^{\ast 2}  \label{20}
\end{equation}
\begin{equation}
\xi ^{\ast 4}=\frac{eB}{4c}\,\xi ^{1}+\frac{1}{2}\,\xi ^{4}=\frac{eB}{2c}%
\,\xi ^{\ast 1}.  \label{21}
\end{equation}
From (\ref{21}) we see that the reduced phase-space is two-dimensional with
the usual canonical commutation relations between $\xi ^{\ast 1}$ and $\xi
^{\ast 2}.$ The Hamiltonian expressed in terms of these two physical
variables can be used as a starting point for the quantization of the
system. In general, the quantization is local on constraint surface, but
this case represents an exception.

\section{Conclusion}

In this paper we have presented a method for projecting the global
coordinates of an Euclidean configuration (phase) space of a system with
holonomic or second class constraints onto the physical surface. The
projection gives free local coordinates on the surface and one can express
the dynamics of the system and the observables in terms of them. The
projection is achieved by a local projector that, in the phase space, is the
matrix product between the symplectic form and the Dirac matrix and its 
trace gives the number of the degrees of freedom of the system.
When the model presents gauge symmetries represented by
first class constraints, one has to fix firstly these symmetries. It would
be interesting to investigate further the possibility of encoding global
information about the constraint surface in the theory and to analyze the
extension of the method to configuration (phase) spaces with more complicate
topology.


\begin{thebibliography}{99}
\bibitem{ht}  M. Henneaux and C. Teitelboim, 
Quantization Of Gauge Systems, Princeton University Press, 1992

\bibitem{cma}  C.~M.~Do Amaral, 
Nuovo Cim.\ B \textbf{25}, 817 (1975). 

\bibitem{pp}  P.~Pitanga and C.~M.~do Amaral, 
Nuovo Cim.\ A \textbf{103}, 1529 (1990). 

\bibitem{qe}  M.~A.~Santos, J.~C.~de Mello and P.~Pitanga, 
Z.\ Phys.\ C \textbf{55}, 271 (1992). 

\bibitem{cl}  M.A. Santos, J.C. de Mello and P. Pitanga, Braz. J.
Phys. \textbf{23}, 214 (1993). 

\bibitem{mj}  M.~A.~Santos and J.~A.~Helayel-Neto, 
hep-th/9905065. 

\bibitem{alv}  L.~R.~Manssur, A.~L.~Nogueira and M.~A.~Santos, 
hep-th/0005214. 

\bibitem{mmi1}  M.~A.~De Andrade, M.~A.~Santos and I.~V.~Vancea, 
JHEP \textbf{0106}, 026 (2001) [hep-th/0104154]. 

\bibitem{dirac}  P.A.M. Dirac, Lectures on Quantum Mechanics, Belfer
Graduate School of Science, New York: Yeshiva University 1964.

\bibitem{suss}  D.~Bigatti and L.~Susskind, 
Phys.\ Rev.\ D \textbf{62}, 066004 (2000) [hep-th/9908056]. 

\bibitem{Kim-Oh}  W.~T.~Kim and J.~J.~Oh, 
Mod.\ Phys.\ Lett.\ A \textbf{15}, 1597 (2000) [hep-th/9911085]. 

\bibitem{sak} J.~J.~Sakurai, Advanced Quantum Mechanics, Addison-Wesley, 1967

\end{thebibliography}
\end{document}